# Topological Critical Point and Resistivity Anomaly in HfTe$_5$

L. X. Zhao[1], X. C. Huang[1], Y. J. Long[1], D. Chen[1], H. Liang[1], Z. H. Yang[1], M. Q. Xue[1], Z. A. Ren[1,2], H. M. Weng[1,2,†], Z. Fang[1,2,3], X. Dai[1,2,3], and G. F. Chen[1,2,3,†]

[1] *Institute of Physics and Beijing National Laboratory for Condensed Matter Physics, Chinese Academy of Sciences, Beijing 100190, China*

[2] *Collaborative Innovation Center of Quantum Matter, Beijing 100190, China*

[3] *School of Physics, University of Chinese Academy of Sciences, Beijing 100049, China*

## Abstract

There is a long-standing confusion concerning the physical origin of the anomalous resistivity peak in transition metal pentatelluride HfTe$_5$. Several mechanisms, like the formation of charge density wave or polaron, have been proposed, but so far no conclusive evidence has been presented. In this work, we investigate the unusual temperature dependence of magneto-transport properties in HfTe$_5$. We find that a three dimensional topological Dirac semimetal state emerges only at around $T_p$ (at which the resistivity shows a pronounced peak), as manifested by a large negative magnetoresistance. This accidental Dirac semimetal state mediates the topological quantum phase transition between the two distinct weak and strong topological insulator phases in HfTe$_5$. Our work not only provides the first evidence of a temperature-induced critical topological phase transition in HfTe$_5$, but also gives a reasonable explanation on the long-lasting question.

The discovery of both two-dimensional (2D) and three-dimensional (3D) topological insulators (TIs) has attracted much attention recent years due to their rich physics and promising applications in electronic and spintronic devices [1-4]. Recently, many of topological phases of matter, such as topological crystalline insulators [5-8], topological Kondo insulators [9-12], topological Dirac [13-17] and Weyl [18-25] semimetals (DSMs/WSMs), have been predicted theoretically and realized experimentally. These exotic topological quantum phases could also be tuned or induced by applying pressure, varying temperature or chemical substitution, etc. [26-32]. For example, by fine tuning the chemical composition to the critical point of the topological phase transition between a normal insulator and a strong TI, the bulk band gap goes to zero at topological critical point (TCP) and the bulk band structure could be described by a 3D Dirac fermion state [30,32].

Quasi-two-dimensional (Q2D) layered compound $HfTe_5$, previously known as thermoelectric materials, has stimulated extensive investigating interests for the pronounced resistivity peak occurred at around $T_P$ = 50 - 90 K [33-35]. However, the origin of the mysterious transport feature has been an unsolved question for a long time [33-37]. Recent theoretical calculations predicted that single-layer $HfTe_5$ is a large gap quantum spin Hall insulator, and the bulk $HfTe_5$ with stacking of many layers locates in the vicinity of a transition between strong and weak TI [38]. This predication has renewed the interest in exploring its topological characters and unveiling the mystery of the origin of the anomalous resistivity peak. In this work, we study the temperature dependence of magneto-transport properties in $HfTe_5$, and find that the observed anomalous resistivity peak is intimately tied to a TCP. We show that, at around $T_p$, magnetoresistance (MR) exhibits an archetypal weak anti-localization (WAL) character in the perpendicular current and magnetic fields. When the magnetic field is rotated to parallel to the current, a chiral anomaly resulted negative MR [39-44] is observed. Above or below $T_p$, however, the negative MR is suppressed gradually. All these results demonstrate clearly that, at around $T_p$, a 3D topological DSM state emerges, which mediates the topological quantum phase transition between the two distinct weak and strong TI phases in HfTe$_5$. By approaching to TCP, the bulk band gap goes to zero and leads to a pronounced resistivity peak. The present finding provides a new perspective in further exploring the intriguing DSM/WSM states and gives a reasonable interpretation for the long-standing confusion of the pronounced resistivity peak in HfTe$_5$.

Single crystals of HfTe$_5$ were grown by chemical vapor transport. Stoichiometric amounts of Hf (powder, 3N, Zr nominal 3%) and Te (powder, 5N) were sealed in a quartz ampoule with iodine (7 mg/mL) and placed in a two-zone furnace. Typical temperature gradient from 500 °C to 400 °C was applied. After one month, long ribbon-shaped single crystals were obtained. Figure 1(a) shows the crystal structure of HfTe$_5$, trigonal prismatic HfTe$_3$ chains run along the $a$-axis, and the largest natural facet determined by X-ray diffraction is (0 1 0) plane which stack along the $b$-axis. Theoretical calculations suggested that the interlayer interaction dominated by van der Waals (vdW) bonding is as weak as that of graphite [38]. Hence, to some extent, this system may be regarded as a Q2D electron system due to the weak interlayer coupling.

The MR was measured with the four-point probe method in a Quantum Design PPMS, and the Hall coefficient measurement was done using a five-probe technique. Electrical contact was made using Au wires bonded to the crystal with Au paint. For MR (or Hall resistivity) measurements, any Hall (or resistive) voltages due to misalignment of the voltage leads could be corrected by reversing the direction of the magnetic field. Figure 1(b) presents a typical temperature dependent resistivity curve $\rho(T)$ in zero magnetic field, which shows a metallic behavior down to $\sim$ 175 K. However, with further decreasing temperature, $\rho(T)$ increases and reaches a maximum at around $T_p \sim$ 65 K. Below 65 K, we observed the metallic-like behavior again with a nearly saturated resistivity at 2 K. This exotic phenomenon is investigated at various magnetic fields and shown in the inset of Fig. 1(b). The applied magnetic fields significantly increase the resistivity and stimulate a crossover from metallic to insulating behavior which may be related to a distortion of the Fermi surface, but the peak is still alive. In order to give more insight into this anomalous peak, following MR measurements are carried out at three typical temperatures ($T$ = 2, 65 and 100 K) as marked in the main panel in Fig. 1(b).

As shown in Fig. 1(c) to 1(f), at $T$ = 2, 65 and 100 K, the MR is measured at different angles, $\theta$, between the electric current and the magnetic field. The insets zoom in on the lower MR parts and depict the correspondingly measurement configurations. At $T$ = 2 K [ Fig. 1(c)], the MR exhibits obvious Shubnikov-de Haas (SdH) oscillations with a very low frequency of $\sim$1 T in the low field range, which result from the depopulation of the Landau levels N = 3, 2 and 1, suggesting the existence of a very small concentration of light carriers. Beyond this quantum limit ($B \sim 2$ T), the MR increases parabolically with B and reaches up to 4000% at 9 T. We also find that the quantum limit depends on the tiltt angle $\theta$, and the MR is suppressed remarkably as $\theta$ is increased. We didn't observe a negative MR at $\theta = 90°$. In contrast to the case at $T = 2$ K, in the inset of Fig. 1(d), a clear ngative MR at $T = 65$ K is detected in the parallel electric and magnetic fields, hinting at the presence of the chiral anomaly of Weyl fermions. Remarkably, the MR decreases monotonously and tends to saturation with increasing magnetic field, indicating the high quality of the sample. Furthermore, we noted that, the MR curve appears to have a broad cusp-like depression, as shown in the main panel of Fig. 1d, which is corresponding to so-called WAL observed commonly in graphene and topological insulators [45-47], in which Dirac fermions dominate the magneto-electric transport. Measurements are also implemented by tilting the magnetic field with respect to the (010) facet but keeping the field perpendicular to the current [Fig. 1(e)]. As expected, the WAL behavior is free from effect, but no trace of negative MR phenomenon has been detected. This adds evidence that the negative MR origins from the chiral term ***E·B***. When the temperature increases to 100 K [Fig. 1(f)], both the negative MR and the WAL disappeared. Namely, the WSM state is destroyed by further increasing the temperature. The whole process presented in Fig. 1(c) to 1(f) can be interpreted as a temperature-induced changing of the topological band structure of $HfTe_5$, which generates non-trival Dirac-like linear dispersion at around $T_p \sim 65$ K. The extrinsic magnetic field breaks the time-reversal symmetry and degenerates the Dirac node to a pair of Weyl nodes. In the following we will discuss in detail the unusually evolution of the negative MR and WAL behaviors when the magnetic field is parallel and perpendicular to the current.

Figure 2(a) and 2(b) present the MR measured at typical temperatures in parallel electric and magnetic fields. At 2 K, the value of MR is positive in the whole range of the applied magnetic fields. Remarkably, there is a negative trend with increasing temperature, and eventually the MR fall to negative 30 per cent at T = 60 K and H = 9 T. This response to temperature change is opposite to that observed in the 3D DSMs/WSMs, such as TaAs, $Na_3Bi$, and $Bi_{1-x}Sb_x$ [21,41,44]. Upon further increasing temperature, the negative MR was gradually suppressed, and the positive MR appeared again above 100 K. When a perpendicular magnetic field is applied, as shown in Fig. 2(c) and 2(d), however, the WAL effct is less affected by lowing temperature, indicating that the topological surface states are remarkably robust at low temperature region. Above $T_p$, as the temperature increases, the cusps are broadened and finally disappear due to the continuous decrease of the phase coherent length. Overall, one can attribute these unique phenomena to the formation of a 3D DSM state at around $T_p$, which has been proposed to appear in the critical point of topological quantun phase transition (TQPT) between two different topological phases through accidental band crossing [26]. For example, by fine tuning the chemical composition in $BiTl(S_{1-\delta}Se_\delta)_2$, a 3D DSM state has been relaized at the critical point of the TQPT [30].

The effects of the negative MR and WAL on transport properties of WSM can be described in a general formula [21,44]:

$$\sigma(B) = (\sigma_0 + C_W B^2) \cdot \sigma_{WAL} + \sigma_N, \tag{1}$$

where,

$$\sigma_{WAL} = \sigma_0 + a\sqrt{B}, \tag{2}$$

and

$$\sigma_N^{-1} = \sigma_0^{-1} + A \cdot B^2. \tag{3}$$

$\sigma_0$ is the zero field conductivity, and $C_W$ is a positive parameter which originates from the topological $\boldsymbol{E}\cdot\boldsymbol{B}$. Such a topological term will generate chiral current in the non-orthogonal magnetic and electric fields. $\sigma_{WAL}$ and $\sigma_N$ are those from WAL effect and normal non-linear band structures around the Fermi level, respectively. At $\theta = 90°$, the contribution from the $\sigma_{WAL}$ term can be considered as a constant $\sigma^o_{WAL}$, since the samples are thin ribbon-like quasi-two-dimensional single crystals. And $C_W$ is a nonzero value. Then, in the weak field region, the chiral conductivity derived from Eq. (1) is expressed by [44]:

$$\sigma^{chiral} = (\sigma_0 + C_W B^2) \cdot \sigma^0_{WAL}. \tag{4}$$

In contrast, at $\theta = 0°$, the topological $\boldsymbol{E}\cdot\boldsymbol{B}$ term is zero, and the contribution from $C_W$ vanished. However, the effect of WAL on the conductivity can't be neglected. As a consequence, Eq. (1) is rewritten as:

$$\sigma(B) = (\sigma_0 + a\sqrt{B}) + (\rho_0 + A \cdot B^2)^{-1}. \tag{5}$$

The measured negative MR ($\theta = 90°$) and WAL ($\theta = 0°$) data at $T = 60$ K are fitted with Eq. (4) and (5), respectively. As shown in Fig. 2(c) and 2(d), the black circles represent the experimental results and the red lines depict the theoretical fittings. The fittings show excellent agreement between the experimental data and theoretical curves with the parameters of $\sigma_0 = 338.4\ \Omega^{-1}\text{cm}^{-1}$, $C_W = 3.04\ \text{T}^{-2}$, $A = 1.38 \times 10^5\ \Omega \cdot \text{cm} \text{T}^{-2}$ and $\rho_0 = 163\ \Omega \cdot \text{cm}$. This confirmed that the negative MR and WAL at around $T_p$ origined from the chiral Weyl fermions, signalling a temperature tuned DSM state in HfTe$_5$.

Magnetic field dependent Hall resistivity $\rho_{xy}$ of HfTe$_5$ at different temperatures is studied and presented in Fig. (3). As shown in Fig. 3(a), at high temperatures, the positive slopes of Hall resistivity indicate that the holes dominate the main transport process. With the temperature down to 80 K, however, both the slopes and the values of Hall resistivity change signs in low fields, implying the carriers dominating the conduction mechanism transformed to electron-type. All these are consistent with multiple hole- and electron-like carriers as observed in TaAs [21]. For a simple two-carrier model, the Hall conductivity $\sigma_{xy}$ is expressed as [48-50]:

$$\sigma_{xy} = \left[ a_1 n_1 \mu_1^2 \frac{1}{1+(\mu_1 B)^2} - a_2 n_2 \mu_2^2 \frac{1}{1+(\mu_2 B)^2} \right] eB, \tag{6}$$

where $n_1$ ($n_2$) and $\mu_1$ ($\mu_2$) represent the concentration and mobility for two kinds of carriers, respectively. We should note that the "two kinds of carriers" may not only represent the coexistence of electrons and holes, but also signal the possibility of the existence of two kinds of electrons (or holes) with distinct dispersion relations. This resorts to the sign of $a_1 n_1$ and $a_2 n_2$, in which $a_1 = -1$ (or $+1$) correspond to electron-carriers (or hole-carriers), and $a_2$ = -1 (or +1) correspond to hole-carriers (or electron-carriers), respectively. Figure 3(b) presents the Hall conductivity $\sigma_{xy}$ (solid lines) and the nicely fitting results (dotted lines) at several temperatures. Using the fitting parameters, we extracted the Hall mobility and carrier concentrations. As shown in Fig. 3(c), the two kinds of Hall mobility exhibit distinct behaviors with respect to temperatures. $\mu_2$ is nearly constant in a wide temperature range from low to high temperatures. However, with decreasing temperature, especially below 65 K, $\mu_1$ increased dramatically and reached up to $2.4 \times 10^5$ cm$^2$V$^{-1}$s$^{-1}$ at 2 K. The high carrier mobility is one of hallmarks of Dirac or Weyl fermions [21, 51]. The unusual behavior around 65 K can be further understood through analyzing the carrier concentrations. Figure 3(d) can be intuitively divided into four different parts. As shown, both of the two kinds of carriers are hole-type at higher temperatures, and the concentrations decrease remarkably with decreasing temperature. At around 110 K, the carriers relate to $\mu_2$ changed from hole-type to electron-type, which corresponds to the metal-insulator transition in resistivity. Notably, the carriers relate to $\mu_1$ changed their type near 65 K, signaling the temperature tuned changing of the Fermi surfaces at this temperature. The inset of Fig. 3(c) shows the temperature dependence of Hall coefficient $R_H$ calculated at $B$ = 1.1 T. It is interesting to note that, $R_H$ sharply reverses its sign at around $T_P \sim$ 65 K, which is very different from that of other multiple band systems, in which $R_H$ changes sign gradually when the dominant carrier type switches [21, 51]. This observation confirmed further the occurrence of the band recombination due to a topological phase transition in HfTe$_5$.

In Ref. [38], the single layer of $ZrTe_5$ or $HfTe_5$ has been proposed to be a 2D TI with large band gap. Therefore, the 3D $HfTe_5$ can be looked as the stacking of the 2D TIs along $b$-axis with quite weak van der Waals interaction. The band dispersion along $b$-axis might lead to band inversion if the interlayer coupling is strong enough. This will lead to a TQPT from 3D weak TI to strong TI. The first-principles calculation indicates that both $ZrTe_5$ and $HfTe_5$ are at the vicinity of such TCP. [38] In Fig. 4(a), there is no band inversion along layer stacking direction $\Gamma$-Z and it is a weak TI. As slightly reducing the interlayer distance, the band gap closes at TCP in Fig. 4(b) and reopens in Fig. 4(c). At the TCP, the Dirac cone like band dispersion at $\Gamma$ can be looked as a class-2 DSM [27]. Combined the theoretical and experimental results, the mystery of the anomalous resistivity in $HfTe_5$ that has perplexed people for more than 30 years has been solved. At high temperatures, the bulk band gap is opened and $HfTe_5$ exhibits semiconducting behavior. With the sample cooling, the bulk band gap goes to zero at a critical temperature $T_p$ and whereafter the surface state dominates the electronic transport properties, which thus leads to a pronounced resistivity peak at around $T_p$.

In summary, we find a temperature-induced change of the topological band structure in $HfTe_5$ as evidenced by magneto-transport measurements. Remarkably, in the critical point of the TQPT between the weak and strong TI phases, a 3D Dirac semimetal state appears through accidental band crossing at the critical temperature $T_p$, manifested by the observation of a large negative MR and WAL effect. By approaching to TCP, the bulk band gap goes to zero and leads to a pronounced resistivity peak. Our work gives a reasonable interpretation of the long-standing confusion of the anomalous resistivity peak in HfTe$_5$.

**Acknowledgements**

The authors wish to thank L. Lu for their fruitful discussions and helpful comments. This work was supported by National Basic Research Program of China 973 Program (Grant No. 2015CB921303, 2011CBA00108 and 2013CB921700), the "Strategic Priority Research Program (B)" of the Chinese Academy of Sciences (Grant No. XDB07020100) and the National Natural Science Foundation of China (No. 11274359, 11422428 and 11421092).

[†]Corresponding authors: gfchen@iphy.ac.cn, hmweng@iphy.ac.cn

# References

1. X. L. Qi and S. C. Zhang, Rev. Mod. Phys. **83**, 1057 (2011).

2. M. Z. Hasan and C. L. Kane, Rev. Mod. Phys. **82**, 3045 (2010).

3. H. M. Weng, R. Yu, X. Hu, X. Dai, and Z. Fang, Adv. Phys. **64**, 227 (2015).

4. H. M. Weng, X. Dai, and Z. Fang, MRS Bulletin **39**, 849 (2015).

5. L. Fu, Phys. Rev. Lett. **106**, 106802 (2011).

6. T. H. Hsieh, H. Lin, J. W. Liu, W. H. Duan, A. Bansil, and L. Fu, Nat. Commun. **3**, 982 (2012).

7. P. Dziawa, B. J. Kowalski, K. Dybko, R. Buczko, A. Szczerbakow, M. Szot, E. Łusakowska, T. Balasubramanian, B. M. Wojek, M. H. Berntsen, O. Tjernberg, and T. Story, Nat. Mater. **11**, 1023 (2012).

8.  Y. Tanaka, Z. Ren, T. Sato, K. Nakayama, S. Souma, T. Takahashi, K. Segawa, and Y. Ando, Nat. Phys. **8**, 800 (2012).

9.  M. Dzero, K. Sun, V. Galitski, and P. Coleman, Phys. Rev. Lett. **104**, 106408 (2010).

10. F. Lu, J. Z. Zhao, H. M. Weng, Z. Fang, and X. Dai, Phys. Rev. Lett. **110**, 096401 (2013)

11. H. M. Weng, J. Z. Zhao, Z. J. Wang, Z. Fang, and X. Dai, Phys. Rev. Lett. **112**, 016403 (2014).

12. N. Xu, P. K. Biswas, J. H. Dil1, R. S. Dhaka, G. Landolt1, S. Muff1, C. E. Matt, X. Shi1, N. C. Plumb, M. Radovic, E. Pomjakushina, K. Conder, A. Amato, S. V. Borisenko, R. Yu, H. M. Weng, Z. Fang, X. Dai, J. Mesot, H. Ding, and M. Shi, Nat. Commun. **5**, 4566 (2014).

13. S. M. Young, S. Zaheer, J. C. Y. Teo, C. L. Kane, E. J. Mele, and A. M. Rappe, Phys. Rev. Lett. **108**, 140405 (2012).

14. Z. J. Wang, Y. Sun, X. Q. Chen, C. Franchini, G. Xu, H. M. Weng, X. Dai, and Z. Fang, Phys. Rev. B **85,** 195320 (2012).

15. Z. J. Wang, H. M. Weng, Q. S. Wu, X. Dai, and Z. Fang, Phys. Rev. B **88**, 125427 (2013).

16. Z. K. Liu, B. Zhou, Y. Zhang, Z. J. Wang, H. M. Weng, D. Prabhakaran, S.-K. Mo, Z. X. Shen, Z. Fang, X. Dai, Z. Hussain, and Y. L. Chen, Science **343**, 864 (2014).

17. Z. K. Liu, J. Jiang, B. Zhou, Z. J. Wang, Y. Zhang, H. M. Weng, D. Prabhakaran,

S-K. Mo, H. Peng, P. Dudin, T. Kim, M. Hoesch, Z. Fang, X. Dai, Z. X. Shen, D. L. Feng, Z. Hussain, and Y. L. Chen, Nat. Mater. **13**, 677 (2014)

18. H. M. Weng, C. Fang, Z. Fang, B. Andrei Bernevig, and X. Dai, Phys. Rev. X **5,** 011029 (2015).

19. X. G. Wan, A. M. Turner, A. Vishwanath, and S. Y. Savrasov, Phys. Rev. B **83,** 205101 (2011).

20. G. Xu, H. M. Weng, Z. J. Wang, X. Dai, and Z. Fang, Phys. Rev. Lett. **107,** 186806 (2011).

21. X. C. Huang, L. X. Zhao, Y. J. Long, P. P. Wang, D. Chen, Z. H. Yang, H. Liang, M. Q. Xue, H. M. Weng, Z. Fang, X. Dai, and G. F. Chen, Phys. Rev. X **5,** 031023 (2015).

22. B. Q. Lv, H. M. Weng, B. B. Fu, X. P. Wang, H. Miao, J. Ma, P. Richard, X. C. Huang, L. X. Zhao, G. F. Chen, Z. Fang, X. Dai, T. Qian, and H. Ding, Phys. Rev. X **5**, 011029 (2015).

23. B. Q. Lv, N. Xu, H. M. Weng, J. Z. Ma, P. Richard, X. C. Huang, L. X. Zhao, G. F. Chen, C. E. Matt, F. Bisti, V. N. Strocov, J. Mesot, Z. Fang, X. Dai, T. Qian, M. Shi, and H. Ding, Nat. Phys. **11**, 724 (2015).

24. S. Y. Xu, I. Belopolski1, N. Alidoust, M. Neupane, G. Bian, C. L. Zhang, R. Sankar, G. Q. Chang, Z. J. Yuan, C. C. Lee, S. M. Huang, H. Zheng, J. Ma, D. S. Sanchez, B. K. Wang, A. Bansil, F. C. Chou, P. P. Shibayev, H. Lin, S. Jia, and M. Zahid Hasan, Science **349**, 613 (2015).

25. L. Lu, Z. Y. Wang, D. X. Ye, L. X. Ran, L. Fu, J. D. Joannopoulos, and M.

Soljačić, Science **349**, 622 (2015).

26. S. Murakami, Phys. E **43**, 748 (2011).

27. B. J. Yang and N. Nagaosa, Nat. Commun. **5**, 4898 (2014).

28. M. S. Bahramy, B. J. Yang, R. Arita, and N. Nagaosa, Nat. Commun. **3**, 679 (2012).

29. X. X. Xi, C. L. Ma, Z. X. Liu, Z. Q. Chen, W. Ku, H. Berger, C. Martin, D. B. Tanner, and G. L. Carr, Phys. Rev. Lett. **111**, 155701 (2013).

30. S. Y. Xu, Y. Xia, L. A. Wray, S. Jia, F. Meier, J. H. Di, J. Osterwalder, B. Slomski, A. Bansil, H. Lin, R. J. Cava, and M. Z. Hasan, Science **332**, 560 (2011).

31. T. Sato, K. Segawa, K. Kosaka, S. Souma, K. Nakayama, K. Eto, T. Minami, Y. Ando, and T. Takahashi, Nat. Phys. **7**, 840 (2011).

32. M. Brahlek, N. Bansal, N. Koirala, S. Y. Xu, M. Neupane, C. Liu, M. Zahid Hasan, and S. Oh, Phys. Rev. Lett. **109**, 186403 (2012).

33. T. J. Wieting, D. U. Gubser, S. A. Wolf, and F. Levy, Bull. Am. Phys. Soc. **25**, 340 (1980).

34. F. J. DiSalvo, R. M. Fleming, and J. V. Waszczak, Bull. Am. Phys. Soc. **26**, 449 (1981).

35. F. J. DiSalvo, R. M. Fleming, and J. V. Waszczak, Phys. Rev. B **24**, 2935 (1981).

36. S. Okada1, T. Sambongi1, M. Ido1, Y. Tazuke1, R. Aoki, and O. Fujita, J. Phys. Soc. Jpn. **51**, 460 (1982).

37. D. N. McIlroy, S. Moore, D. Q. Zhang, J. Wharton, B. Kempton, R. Littleton, M. Wilson, T. M. Tritt, and C. G. Olson, J. Phys.: Condens. Matter **16**, L359 (2004).

38. H. M. Weng, X. Dai, and Z. Fang, Phys. Rev. X **4**, 011002 (2014).

39. P. Hosur and X. L. Qi, C. R. Phys. **14**, 857 (2013).

40. D. T. Son and B. Z. Spivak, Phys. Rev. B **88**, 104412 (2013).

41. J. Xiong, S. K. Kushwaha, T. Liang, J. W. Krizan, M. Hirschberger, W. D. Wang, R. J. Cava, N. P. Ong, Science **350**, 413 (2015).

42. S. A. Parameswaran, T. Grover, D. A. Abanin, D. A. Pesin, and A. Vishwanath, Phys. Rev. X **4**, 031035 (2014).

43. A. C. Potter, I. Kimchi, and A. Vishwanath, Nat. Commun. **5**, 5161 (2014).

44. H. J. Kim, K. S. Kim, J. F. Wang, M. Sasaki, N. Satoh, A. Ohnishi, M. Kitaura, M. Yang, and L. Li, Phys. Rev. Lett. **111,** 246603 (2013).

45. X. S. Wu, X. B. Li, Z. M. Song, C. Berger, and W. A. de Heer, Phys. Rev. Lett. **98,** 136801 (2007).

46. M. H. Liu, J. S. Zhang, C. Z. Chang, Z. C. Zhang, X. Feng, K. Li, K. He, L. L. Wang, X. Chen, X. Dai, Z. Fang, Q. K. Xue, X. C. Ma, and Y. Y. Wang, Phys. Rev. Lett. **108,** 036805 (2012).

47. S. Ishiwata, Y. Shiomi, J. S. Lee M. S. Bahramy, T. Suzuki, M. Uchida, R. Arita, Y. Taguchi, and Y. Tokura, Nat. Mater. **12,** 512 (2013).

48. H. Takahashi, R. Okazaki, Y. Yasui, and I. Terasaki, Phys. Rev. B **84,** 205215 (2011).

49. B. Xia, P. Ren, A. Sulaev, P. Liu, S. Q. Shen, and L. Wang, Phys. Rev. B **87,** 085442 (2013).

50. L.Fu and C. L. Kane, Phys. Rev. B **76**, 045302 (2007).

51. T. Liang, Q. Gibson, M. N. Ali, M. H. Liu, R. J. Cava, and N. P. Ong, Nat. Mater. **14**, 280 (2015).

**Figure Captions**

FIG. 1. Crystal structure and transport results of $HfTe_5$. (a) Crystal structure of $HfTe_5$ with Cmcm ($D_{2h}^{17}$) space group. The yellow trigons highlight $HfTe_3$ chain which run along the a-axis and linked via zigzag chains of Te atoms denoted by red dashed line. The main interaction between layers is van der Waals force. (b) Main panel: temperature dependent resistivity in zero field. Three typical temperature points of T = 2, 65 and 100 K for further magneto-resistivity measurements are marked by red, green and blue color, respectively. Inset panel: temperature dependent resistivity in the field of 1, 3, 5, 7 and 9T. The applied magnetic fields not only significantly increase the resistivity, but also stimulate metallic-insulating transitions at low temperatures in definite strength. (c-f) Magneto-resistance with applied field from perpendicular ($\theta = 0°$) to parallel ($\theta = 90°$) to the electric current at 2, 65 and 100 K, respectively. The insets zoom in on the lower MR parts and depict the correspondingly measurement configurations. $\theta$ is defined as the angle between the magnetic field and the electric field. The negative MR and the WAL behavior appear at around 65 K – the temperature of the anomalous resistivity peak. (e) Angular dependent MR at 65 K with keeping magnetic field perpendicular to electric current. The inset shows the measurement configuration. $\Phi$ is defined as the angle between B and B'. The disappeared negative MR implied that it may stem from the topological **_E·B_** term.

FIG. 2. Observation of the Weyl semimetal state under magnetic fields in HfTe$_5$ at around $T_p$. (a, b) Magneto-resistivity at different temperatures with the applied magnetic field parallel to the electric current. (c, d) Magneto-resistivity at different temperatures with the applied magnetic field perpendicular to the electric current. Inset shows MR with expanded scale. (e, f) Negative MR (B//E) and WAL effect (B$\perp$E) at 65 K in weak magnetic fields. The black circles and the red line represent the experimental results and theoretical fitting, respectively. All these signify the appearance of the chiral electrons with Berry's phase of $\pi$.

FIG. 3. Temperature dependence of Hall resistivity, carrier mobility and carrier density for HfTe$_5$. (a) Hall resistivity measured at various temperatures from 2 to 250 K. (b) Magnetic field dependence of $\sigma_{xy}$ at various temperatures. The dotted and the solid lines represent the measuring and the fitting data, respectively. There is a large error in the data fitting for 65 K. (c) Temperature dependence of carrier mobility $\mu_1$, and $\mu_2$ for different carriers deduced by two-carrier model. Inset: The large resistivity peak at zero magnetic field and Hall coefficient obtained by fitting the experimental data below 1.1 T. (d) Temperature dependence of carrier density $n_1$ and $n_2$ for different carriers deduced by two-carrier model.

FIG. 4. Topological quantum phase transition driven by interlayer coupling.The parity configuration (+ means even parity and − means odd one) at eight time-reversal invariant momenta (TRIM) in (a) weak TI phase, (b) TCP and (c) strong TI phase,

respecively. The corresponding band dispersion along layer stacking direction Γ-Z is shown in (d)-(f) in ordering of increasing interlayer coupling. In (f) the band crossing (black dotted line) in case without spin-orbit coupling is open when SOC is considered.

**FIG. 1**

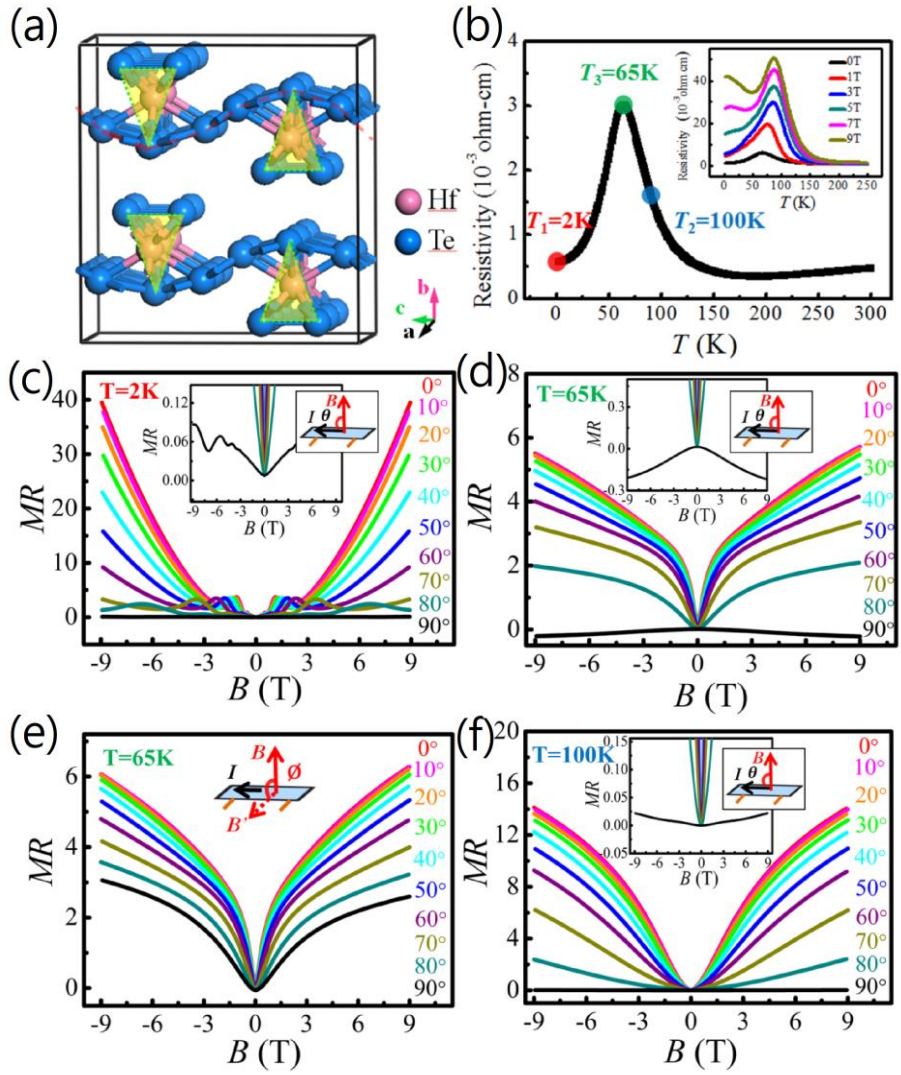

**FIG. 2**

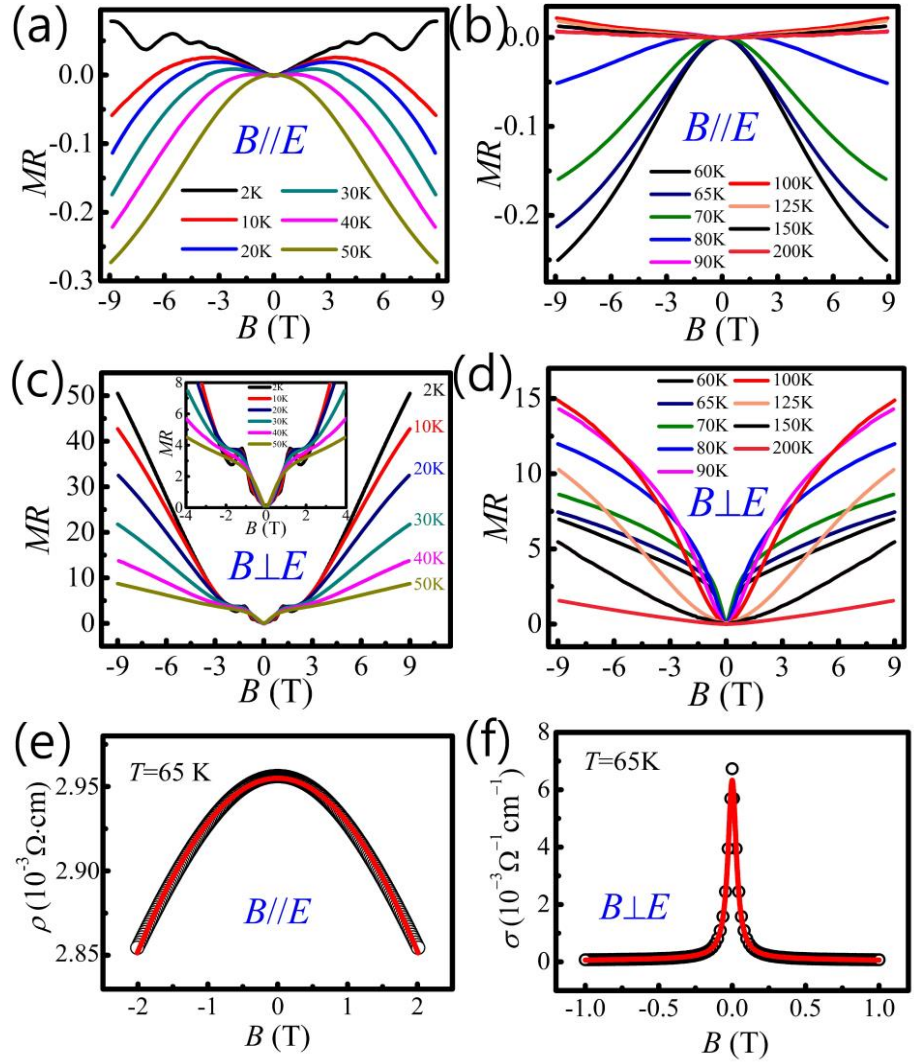

**FIG. 3**

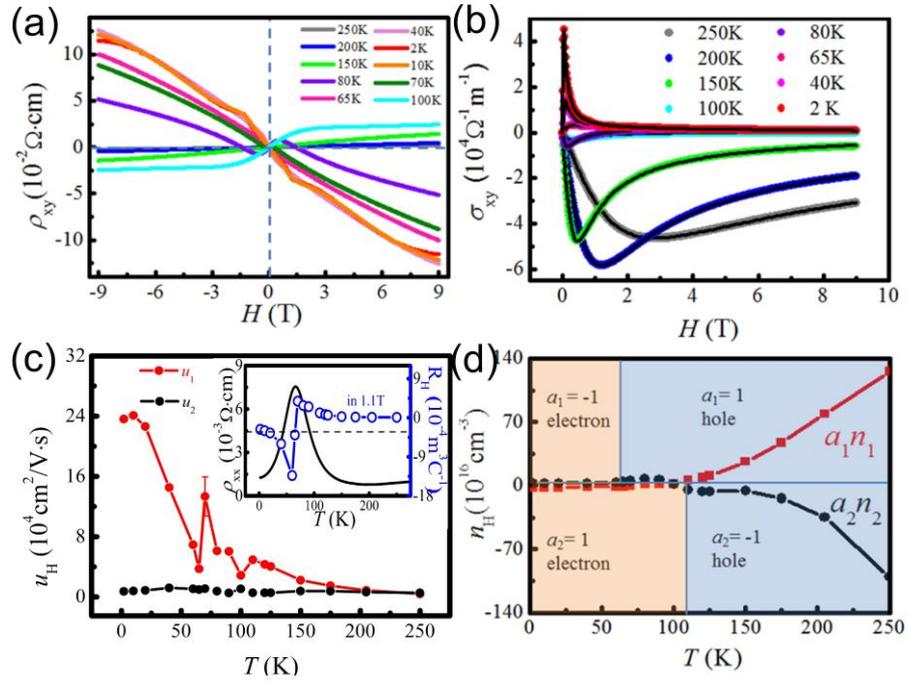

**FIG. 4**

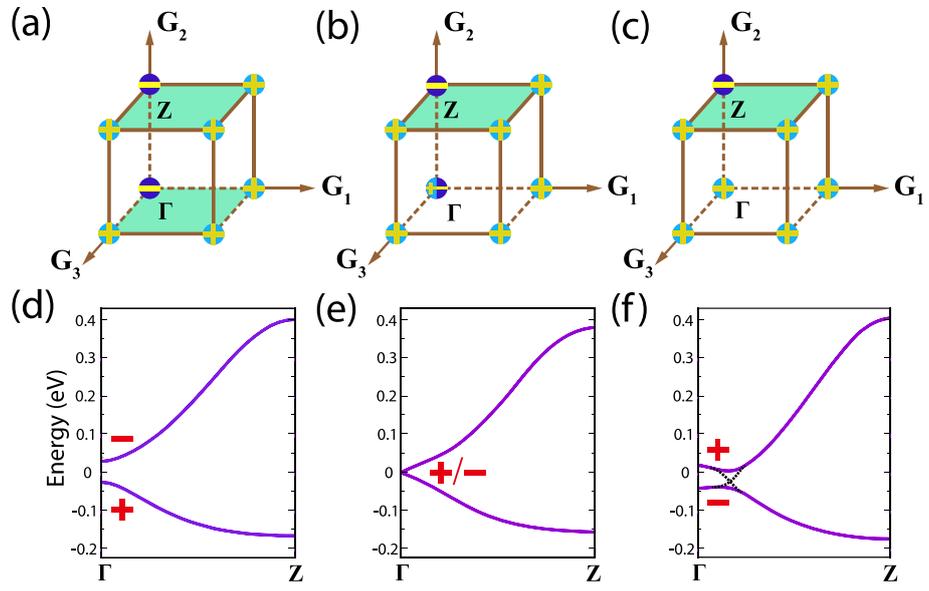